\documentclass[]{emulateapj}

\begin{document}

\title{Spectro-astrometric imaging of molecular gas within protoplanetary disk gaps
\footnote{This work is based on observations collected at the European Southern Observatory 
Very Large Telescope under program ID 179.C-0151.}
}

\author{Klaus M. Pontoppidan\altaffilmark{1,2}}

\author{Geoffrey A. Blake\altaffilmark{1}}

\altaffiltext{1}{California Institute of Technology, Division of Geological and Planetary Sciences, 
MS 150-21, Pasadena, CA 91125}
\altaffiltext{2}{Hubble Fellow, pontoppi@gps.caltech.edu}

\author{Ewine F. van Dishoeck\altaffilmark{3,4}}
\altaffiltext{3}{Leiden Observatory, Leiden University, P.O. Box 9513, NL-2300 RA Leiden, The Netherlands}
\altaffiltext{4}{Max-Planck-Institut f{\"u}r extraterrestrische Physik, P.O. Box 1312, D-85741 Garching, Germany}

\author{Alain Smette\altaffilmark{5}}
\altaffiltext{5}{ESO, Casilla 19001, Santiago 19, Chile}

\author{Michael J. Ireland\altaffilmark{1}}

\author{Joanna Brown\altaffilmark{4,6}}
\altaffiltext{6}{California Institute of Technology, Division of Physics, Mathematics and Astronomy, 
MS 105-24, Pasadena, CA 91125}

\begin{abstract}
We present velocity-resolved spectro-astrometric imaging of the 4.7\,$\mu$m rovibrational lines of CO gas 
in protoplanetary disks using the CRIRES high resolution infrared spectrometer on the Very Large Telescope (VLT).
The method as applied to three disks with known dust gaps or inner holes out to 4-45\,AU (SR 21, HD 135344B and TW Hya)
achieves an unprecedented spatial resolution of $0.1-0.5$\,AU. 
While one possible gap formation mechanism is dynamical clearing
by giant planets, other equally good explanations (stellar companions, grain growth, photo-evaporation) exist. One way of distinguishing
between different scenarios is the presence and distribution of gas inside the dust gaps.
Keplerian disk models are fit to the spectro-astrometric position-velocity curves to derive
geometrical parameters of the molecular gas.
We determine the position angles and inclinations of the inner disks
with accuracies as good as 1-2\degr, as well as the radial extent of the gas emission.
Molecular gas is detected well inside the
dust gaps in all three disks. The gas emission extends to within a radius of 0.5\,AU for HD 135344B and to 0.1\,AU for TW Hya,
supporting partial clearing by a $< 1-10\,M_{\rm Jup}$ planetary body as the cause of the observed dust gaps, or
removal of the dust by extensive grain coagulation and planetesimal formation. The molecular gas emission in SR 21 appears to
be truncated within $\sim 7\,$AU, which may be caused by complete dynamical clearing by a more massive companion.  
We find a smaller inclination angle of the inner disk of TW Hya than that determined for the outer disk, suggestive of a disk warp.
We also detect significant azimuthal asymmetries in the SR 21 and HD 135344B inner disks.
\end{abstract}

\keywords{techniques: high angular resolution -- planetary systems: protoplanetary disks -- stars: individual (\object{EM* SR 21}, \object{HD 135344B}, \object{TW Hya}) -- ISM: molecules}

\section{Introduction}

The recent identification of a significant class of proto-planetary disks
with gaps and inner holes in their radial dust distribution
has given rise to a debate regarding their origin \citep{Najita07}. Dynamical clearing by young giant planets
is the most popular explanation \citep{Strom89,Skrutskie90}, but other potential explanations exist.
The presence of an inner region in the disk that is optically thin to continuum photons from near-infrared to millimeter wavelengths
is not sufficient evidence for a newly formed planet, but simply indicates a deficit of small dust grains. 

A stellar companion orbiting at $\ga$1\,AU, but too close to the primary to
be resolved with direct imaging, would clear out a large hole completely \citep{Artymowicz94}, 
such as is the case for CoKu Tau/4 \citep{Ireland08}.
However, it has been demonstrated that a companion body smaller than 
a certain threshold mass will allow some gas and small dust grains to accrete through its orbit \citep{Lubow99, Kley01}, 
specifically if the Roche lobe of the companion is smaller than the disk scale height.  
The threshold mass depends on many parameters, but reported values are between 1-10 Jupiter masses \citep{Lubow99}, 
and the presence of gas inside a dust hole or gap therefore seems to rule out clearing by a stellar companion. 
Photo-evaporation will create holes in the radial gas and dust distribution extending all the way to the stellar surface, 
if the disk is not too massive \citep{Alexander07}. Finally, 
removal of dust opacity by a systemic growth of dust particles to
meter-sized bodies, or even planetesimals, will remove the dust emission, but
preserve the gas content \citep{Dullemond05}. The various scenarios may be distinguished by the radial 
distribution of molecular gas inside the optically thin regions, since some of the mechanisms for gap
formation tend to remove or destroy molecular gas over a specific range of radii, while others do not. 

In this paper we distinguish between ``gaps'', which are radial clearings leaving an inner disk intact, and ``holes'', which
refer to cases where no inner disk is present. We also distinguish between ``dust opacity gaps/holes'' (shortened to dust gaps) 
and ``gas gaps/holes'', each referring
to separate components traced by different sets of observational data. Specifically, in
this paper we explore whether (molecular) gas gaps are present in disks with dust gaps.

Regardless of the mechanism forming actual disk gaps and holes, it also
essential to measure the radial gas surface density to understand the evolution of solids
and the formation of planetesimals.
The dynamics of large particles, in particular of radii 1-100\,cm, are
intimately coupled to the radial profile of the gas surface density.
This is because $10-100\,$cm sized solids tend to migrate radially against the gas 
pressure gradient \citep{Weidenschilling77}. Thus, the
presence of ``pressure bumps'' in the disk may concentrate the densities of such ``pebbles'' and ``boulders'', 
potentially leading to the formation of planetesimals \citep{Johansen07}. Further, gaps in the radial gas 
distribution will block the inwards migration of particles
rich in volatiles, strongly affecting the chemistry of the inner disk \citep{Ciesla06}. These processes 
take place in the planet-forming zones of protoplanetary disks ($<10\,$AU) requiring
imaging of (molecular) gas on 1-10 milli-arcsec scales.

\begin{figure}
\centering
\includegraphics[width=5cm,angle=90]{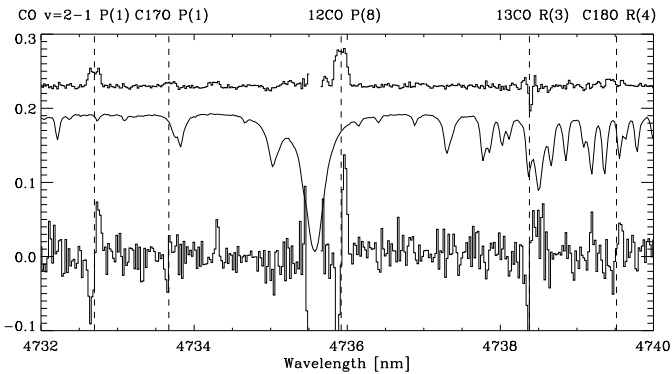}
\caption[]{A small section of the spectral range covered for SR 21 showing
the flux spectrum ({\it top}), the atmospheric transmission spectrum ({\it middle}) and the spectro-astrometric 
signal ({\it bottom}). The units on the vertical axis are in pixels, relevant for the spectro-astrometric signal; the
other two curves are scaled for clarity. The emission lines in the $v=1-0$ rovibrational band (unless otherwise indicated) 
intrinsic to the SR 21 disk are marked with vertical dashed lines.}
\label{SingleLines}
\end{figure}

In this paper we introduce spectro-astrometric imaging of molecular gas in protoplanetary disks as a technique
that allows a spatial resolution better than 1 milli-arcsec ($\sim$ 0.1 AU at the distance of the nearest star forming regions). 
The data not only confirm that molecular gas in Keplerian motion is present well inside the 
dust gaps of three disks, but also measure the distribution (radially and azimuthally) of the emitting gas relative to the dust. 
The disks, TW Hya, SR 21 and HD 135344B, are part of the class of so-called 
``transitional'' or ``cold'' disks, which show dust gaps or holes out to 4, 18 and 45\,AU, respectively \citep{Calvet02,Brown07}, 
at distances of 51, 120 and 84\,pc \citep{Mamajek05, Loinard08, Sylvester96}. 
HD 135344B and SR 21 are examples of disks with dust gaps, since strong near-infrared veiling requires the presence of
an inner disk component at $\lesssim 0.5\,$AU \citep{Brown07}. TW Hya appears to 
have a 4\,AU dust hole \citep{Calvet02}, although recent
near- and mid-infrared interferometry has detected some dust emission within 4 AU \citep{Eisner06,Ratzka07}. 

While HD 135344B and TW Hya both show varying degrees of accretion at relatively low rates 
($5\times 10^{-10}$ to $5\times 10^{-9}\,M_{\odot}\,\rm yr^{-1}$), inferring the 
presence of some gas very close to the central stars ($< 0.1\,$AU) \citep{Muzerolle00,GarciaLopez06}, 
this does not trace the distribution
of gas in the disk at larger radii. No estimate of the accretion rate in SR 21 is available, but the weak 
Pf$\beta$ line observed at 4.65\,$\mu$m suggests that it is low. The mere presence of rovibrational CO lines suggests
that the gas is relatively close to the central star since high temperatures
are required to excite them \citep{Najita03,Salyk07}. However, the CO excitation temperature itself is not 
enough to locate the gas since it is known that
rovibrational lines are often non-thermally excited to large radial distances \citep{Blake04,Goto06}, introducing
significant ambiguities in conclusions based on an assumption of thermalized lines.

\section{Observations}

Traditional infrared imaging cannot resolve the structures in the inner planet-forming regions
of circumstellar disks. Even with modern adaptive-optics assisted telescopes, the very best resolution achieved at 2-5\,$\mu$m
is roughly defined by the airy ring at $1.22 \lambda/D=60-150$ milli-arcseconds for an 8.2\,m telescope, 
corresponding to 6-15 AU at the relevant distance
of 100\,pc \citep{Goto06}. 
Infrared interferometers operating
at 2 $\mu$m can reach resolutions about 20-30 times better, but cannot trace the gas kinematics at 1-10 AU in protoplanetary disks.

Here, we employ spectro-astrometry as a technique for obtaining very high spatial and spectral resolution line
imaging. This technique has been applied at visible wavelengths \citep{Takami01,Porter05}, 
but not yet to longer infrared wavelengths and to molecular lines in protoplanetary disks. 
We demonstrate the feasibility of using spectro-astrometry
for imaging the $\Delta v = 1$ rovibrational band of CO at 4.7\,$\mu$m to a precision $\le$1 milli-arcsec.
The method is based on the fact that the centroid of the image of
a point source can be determined to much higher precision than the
size of the point spread function (PSF), by several orders of magnitude. 
By measuring the centroid (1st order moment) in the spatial direction of a spectrally resolved emission line, 
one can construct position-velocity diagrams, given a certain position angle of the entrance slit of the spectrometer on
the sky.

For this purpose we used the recently commissioned high resolution 
($\lambda/\Delta \lambda \sim 100,000$) infrared spectrometer,
CRIRES, mounted on UT1 of the Very Large Telescope of the European Southern Observatory \citep{Kaufl04}. The 
instrument is fed by the Multi-Application Curvature Adaptive Optics (MACAO) system, making
it ideal for spectro-astrometry. 
The spectro-astrometric technique as applied to a larger suite of CRIRES spectra is described in greater 
detail in an upcoming paper (Pontoppidan et al., in prep.).

We obtained spectra centered at 4.715 \,$\mu$m in the CO $P$-branch 
at 6 different position angles (PA) for each disk. The slit width was 0\farcs2, compared to the diffraction-limited
PSF core of 0\farcs18. The 2-D spectra were processed using
standard reduction techniques for infrared spectroscopy. The formal spatial centroid of the 2-D spectra is
calculated as:

\begin{equation}
X(v) = C \frac{\sum_i (x_i(v)-x_0) F_i(v)}{\sum_i F_i(v)}, \mbox{ [pixels] }
\label{sadef}
\end{equation}
where $x_i(v)-x_0$ is the center of pixel $i$ relative to the continuum centroid at velocity $v$ in the spatial direction and $F_i(v)$ is the flux
in that pixel. The sum is over a given virtual aperture in the spatial direction, in our case $\sim \pm 2\sigma$ of the
PSF, or 400 milli-arcsec. $C$ is a correction factor of order unity that takes into account that not all of the
source flux is enclosed in the virtual aperture. For our choice of aperture, which maximizes the signal-to-noise, 
$C=1.3$. This value was confirmed by measuring the amplitude of the spectro-astrometric signature as a function of virtual 
aperture size. 
Further, the total centroid is diluted by the continuum flux such that the true centroid of the line is
$X_l(v) = X(v) (1+F_c(v)/F_l(v))$, where $F_c/F_l$ is the total continuum-to-line ratio at velocity $v$. 
SR 21 and HD 135344B have $F_c/F_l \sim 5$, lowering the spectro-astrometric sensitivity by a factor 6, while
TW Hya has much smaller continuum dilution with $F_c/F_l \sim 0.7$.

The PAs are paired such that each spectrum has
a counter-spectrum obtained at an angle rotated by 180$\degr$ \citep{Brannigan06}. 
By pairwise subtracting the parallel and anti-parallel spectro-astrometric signals,
artifacts stemming from instrumental effects are removed while any
real signal is retained, i.e. $X(v) = (X_{\rm 0\degr}-X_{\rm 180\degr})/2$.
Ground based 4.7\,$\mu$m spectra are filled with numerous strong absorption lines due to atmospheric 
molecules, including CO itself. We use these strong spectral features to test for the presence of 
artifacts in our spectro-astrometry. For the data presented here, no telluric features
produce false spectro-astrometric signals in excess of the statistical errors after pair subtraction (down to $X(v)\sim 200-500\,\mu$arcsec), 
ruling out the presence of artifacts as discussed in \cite{Brannigan06}. This is illustrated in Figure \ref{SingleLines}, which
shows that the spectro-astrometry is consistent for all the lines intrinsic to the disk, but is not correlated with the
shape of the telluric absorption spectrum. 

\begin{figure*}
\includegraphics[width=17cm]{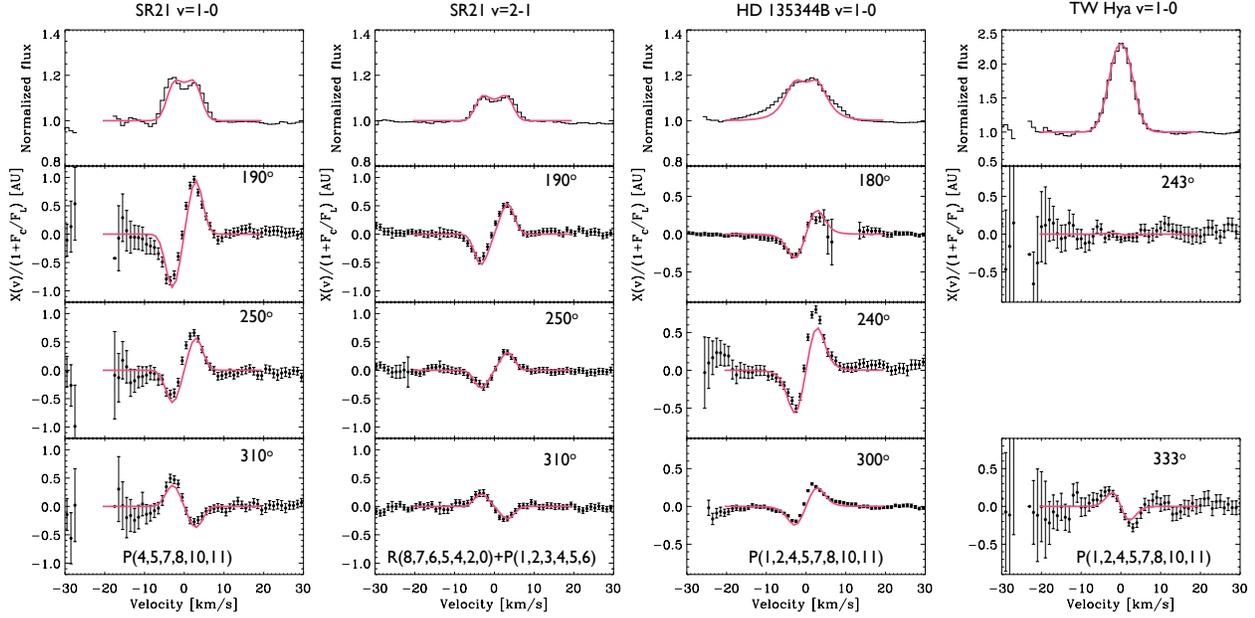}
\caption[]{Top panel: average CO P-branch lines in the $v$=1-0 fundamental rovibrational band at 4.7\,$\mu$m for the 
three cold disks. For SR 21, the $v$=2-1 lines are displayed as well. Bottom panels: Spectro-astrometry for the same
lines as defined in eq. \ref{sadef}; the y-axis offset is for the line+continuum signal with respect to the continuum 
emission centroid. The red curves
show the best fitting disk models. At the bottom of each spectrum the transitions used for the composite are indicated. 
The telluric CO lines are visible as a gap in the spectra. The errors have been propagated from the pixel RMS variations in the
2-D spectrum, under the assumption that the error is dominated by background noise (a good approximation at 4.7\,$\mu$m). }
\label{PV}
\end{figure*}

The CRIRES setting employed covers $\sim$8 CO $v$=1-0 lines and $\sim$13 CO $v$=2-1 lines.
The spectro-astrometric lines from each vibrational band are co-added
to reveal position-velocity diagrams at 3 position angles for each disk (2 for TW Hya), obtained with relative
position angle spacings of 60$\degr$ (90$\degr$ for TW Hya). Care was taken to avoid lines
contaminated by other transitions or foreground absorption. In particular the three lowest $J$ transitions for SR 21 
show an absorption component due to gas in the foreground molecular cloud. While such absorption will
not affect the spectro-astrometry, it will affect the line flux profile. Thus these lines were not
used for the co-added flux spectra and spectro-astrometry.
The flux spectra were obtained by division with an early-type telluric standard 
(HR 6084 for HD 135344B and SR 21; HR 4023 for TW Hya). 

\begin{table}
\centering
\caption{Log of observations}
\begin{tabular}{lllll}
\hline
\hline
Star  & PA & Obs. Date & Int. Time & Spectral Range \\
\hline
SR 21      & 190\degr & 30/8/2007& 32 min& 4.660-4.770\,$\mu$m \\
SR 21      & 250\degr & 30/8/2007& 32 min& 4.660-4.770\,$\mu$m \\
SR 21      & 310\degr & 31/8/2007& 32 min& 4.660-4.770\,$\mu$m \\
HD 135344B & 180\degr & 22/4/2007& 20 min& 4.645-4.755\,$\mu$m \\
HD 135344B & 240\degr & 4/9/2007 & 20 min& 4.660-4.770\,$\mu$m \\
HD 135344B & 300\degr & 5/9/2007 & 20 min& 4.660-4.770\,$\mu$m \\
TW Hya     & 243\degr & 26/4/2007& 40 min& 4.660-4.770\,$\mu$m \\
TW Hya     & 333\degr & 26/4/2007& 40 min& 4.660-4.770\,$\mu$m \\
\hline
\end{tabular}

\label{obs_table}
\end{table}

Ideally, each position angle for a given source should be observed
within a time span of a few days. This is the case for the observations presented, except
for the $\rm PA=180\degr$ of HD135344B which was observed 4 months prior to
the other position angles of that source. The observations are summarized in Table \ref{obs_table}. 

\begin{figure*}
\includegraphics[width=17cm]{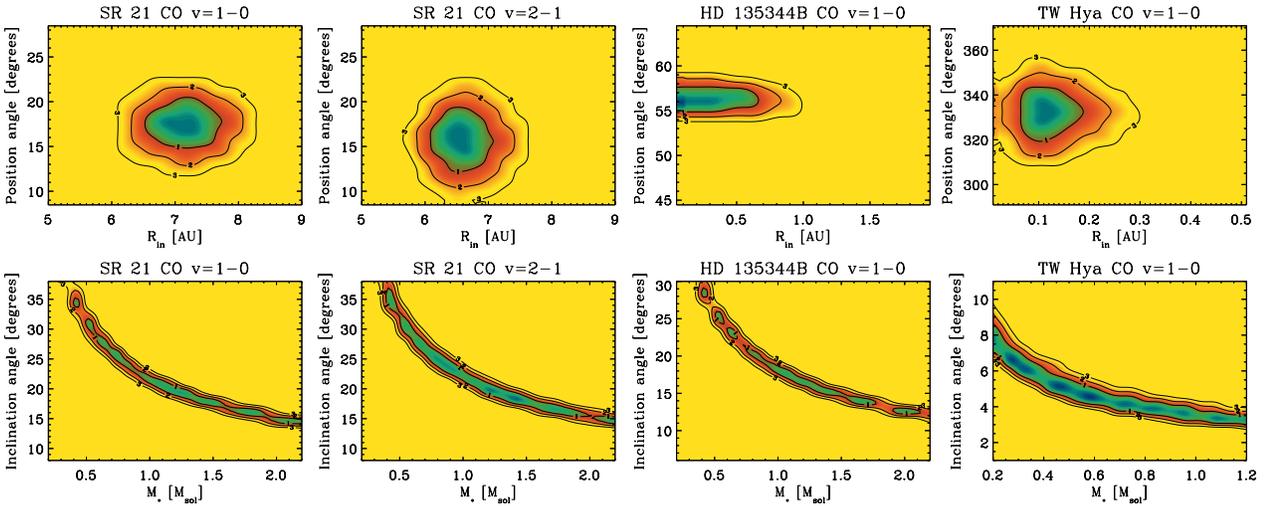}
\caption[]{$\chi^2$ surfaces for the best fitting Keplerian disk models. The contours are at 68, 95 and 99\% confidence levels, corresponding
to 1, 2 and 3$\sigma$.}
\label{Chi2}
\end{figure*}

\section{Results}

\subsection{Keplerian disk models}

An emission line originating in a Keplerian disk is expected to show an antisymmetric position-velocity spectrum, with
the blue- and red-shifted sides of the line offset in opposite directions. 
The CO position-velocity diagrams for SR 21, HD 135344B and TW Hya are shown in Fig. \ref{PV}. 
The CO $\Delta v$=1 lines typically probe molecular gas at temperatures of 500-1200 K \citep{Blake04}. 
It was possible to extract high quality spectro-astrometric signals from both CO $v=$1-0 and CO $v$=2-1
lines in SR 21. All the astrometric CO spectra, with the exception of TW Hya at PA=63$\degr$, show
clear Keplerian signatures. 

In order to infer the physical parameters of the gas emission, a simple model of a Keplerian disk 
is fitted to each set of position-velocity diagrams. The model is constructed as a discrete set of emitting 
rings with a radial temperature dependence determined by a power law: $T=T_{\rm 1\,AU} (R/1\,{\rm AU})^{-q}$, and a constant radial
surface density.
The free parameters in the model are the radius of the innermost ring, the PA of the
major axis of the projection of the disk on the sky, the inclination of the disk and the stellar mass. 
The models were fitted simultaneously to all the spectro-astrometric position angles, as well as 
the line flux spectrum. 

The spectro-astrometric signature is quite sensitive to the PA and the weighted average of the radius of the emitting gas, and these parameters
are generally determined to high accuracy. The stellar mass and disk inclination are degenerate, tracing
a narrow strip of good fits as seen in the $\chi^2$ surfaces shown in Fig. \ref{Chi2}. 
While the spectro-astrometry measures the radius of the weighted average of the line emission, the
constraints of the corresponding line profile add information about the radial extent of the gas.  

For SR 21, the data could be fit with a single ring ($\Delta R/R \ll 1$) at 7-7.5\,AU because the line shape
is double peaked (no emission from large radii) and has steep wings (no emission from small radii).
Attempts at fitting the SR 21 data with a temperature distribution lead to very similar results.
Essentially, it is required that the line surface brightness drops rapidly with increasing radius so as to not contribute to the total
line profile. However, an actual truncation of the gas at some outer radii is not strictly required,  
since it is also possible that it is simply not warm enough to be excited. This scenario constrains the temperature profile to $q\gtrsim 0.4$. 
We consider the more complex model with relatively cold gas extending outwards from an inner cut-off radius 
to be more physical and therefore adopt it, even though a single ring also fits the data.

For HD 135344B and TW Hya, a ring was insufficient to fit both the spectro-astrometry and the 
line profile in detail, suggesting that the line emission comes from a range of radii (see Table \ref{result_table}).
The cause of this is most clear for HD 135344B, which has a line profile with broad wings. 
For TW Hya, we use the temperature profile estimated for the dust using near-infrared interferometry, 
$T_{\rm 1\,AU}=360\,$K and $q=0.4$ \citep{Eisner06}. This value for $q$ corresponds to
optically thin dust in LTE, $q=2/(4+\beta)$, where $\beta=1$ is the dust opacity power law index for small (non-grey) dust grains. 
For HD 135344B, $T_{\rm 1\,AU}=790\,$K and $q=0.4$ provide a 
reasonable fit to the line profile, although the line wings are so broad that a single power law temperature profile 
fails to provide a perfect fit. A shallower temperature
profile with $q<0.4$ over a range of radii may improve the fit, but exploring more complex models requires the inclusion of more
physics, such as fluorescence excitation, in the excitation temperature profile, and is beyond the scope of this paper.  
Further, the line is asymmetric and the blue line wing is stronger than what can be fitted by our simple model.
For SR 21, $T_{\rm 7\,AU}=300\,$K and $q=0.4$ was assumed, relevant for LTE conditions. 
Note, however, that given the strength of the CO $v=2-1$ lines, 
the vibrational ladder is probably non-thermally excited, even if the rotational is not. 

The most obvious result is that the molecular emission originates well within the dust gaps reported in
the literature, and that the radial distribution of the line
emission is not the same for the three disks. SR 21 appears to be devoid of molecular gas within $\sim 5$\,AU, while the molecular gas 
extends to within 1 AU for HD 135344B. 
The molecular gas in TW Hya is found to extend inwards to 0.1\,AU, which is well inside the
4 AU optically thin (at $\lambda \gtrsim 1\,\mu$m) region
reported by \cite{Calvet02}. 

The inclination determined from
the recent Submillimeter Array (SMA)
images (0$\farcs$5 resolution) of the SR 21 disk \citep{Brownthesis07} 
can be used to break the inclination-mass degeneracy in the spectro-astrometry and determine the stellar mass. 
Specifically, an inclination of $20^{\circ}$ gives a stellar mass of
1.1$\pm0.1\,M_{\odot}$ for SR 21. While HD 135344B is viewed too face-on to get a good 
estimate of the inclination from the SMA images,
a very broad assumption on the stellar mass of 1.0-2.0\,$M_{\odot}$, given the spectral type (F4), 
still gives a well-determined inclination of $14\pm 3^{\circ}$. 
Similarly, using a broad range of stellar masses of $0.7\pm0.2\,M_{\odot}$ for TW Hya, the inclination of the inner disk is constrained
to $4.3\pm1.0\degr$.

\begin{table}
\centering
\caption{Best fit model parameters}
\begin{tabular}{llllll}
\hline
\hline
Star  & $R_{\rm in}$ [AU] & $R_{\rm out}$ [AU] & PA & $i$  & $M_*$ [M$_{\odot}$] \\
\hline
SR 21 $v$\,=\,1-0      & 7.6$\pm$0.4     & $>10$     & 16$\pm$3$\degr$ & 22$\pm$4$\degr$\tablenotemark{a} & 0.9$\pm$0.1\tablenotemark{b} \\
SR 21 $v$\,=\,2-1      & 7.0$\pm$0.4     & $>10$     &15$\pm$4$\degr$ & 20$\pm$5$\degr$\tablenotemark{a} & 1.0$\pm$0.1\tablenotemark{b} \\
HD 135344B $v$\,=\,1-0 & 0.3$\pm$0.3     & $>15$     &56$\pm$2$\degr$  & 14$\pm$4$\degr$\tablenotemark{c} & -- \\
TW Hya $v$\,=\,1-0     & 0.11$\pm$0.07   & $>1.5$    &332$\pm$10$\degr$ & 4.3$\pm$1.0$\degr$\tablenotemark{d}& -- \\
\hline
\end{tabular}
\tablenotetext{a}{Assuming $M_*=0.7-1.5\,M_{\odot}$.}
\tablenotetext{b}{Assuming $i=20^{\circ}$ \citep{Brownthesis07}. }
\tablenotetext{c}{Assuming $M_*=1.0-2.0\,M_{\odot}$.}
\tablenotetext{d}{Assuming $M_*=0.5-0.9\,M_{\odot}$.}

\label{result_table}
\end{table}

\subsection{Azimuthal asymmetries}

Because the models presented are azimuthally symmetric and assume isotropic local line emission, 
they produce symmetric (to inversion of the
velocity axis) line profiles and symmetric (to inversion of both the velocity and the position axis)
position-velocity diagrams. However, significant departures from these symmetries are apparent in the data, showing
that the line emission from the inner disks is not azimuthally symmetric about the continuum centroid. 
This is especially apparent in SR 21 CO ($v$=1-0) at PA=250 and 310$\degr$, as well as
in HD 135344 at PA=240$\degr$. The spectro-astrometric technique is thus well-suited for detecting disk asymmetries at a specific
azimuth angle $\psi$. The caveat is that it cannot directly distinguish between an offset in radius in the 
line emission at $\psi$ and an offset in the 
continuum emission in the opposite direction, $\psi + \pi$. We define the asymmetry direction as the azimuth angle at which the line
emission radial offset is positive, $X_{\rm asym}(v) = X(v)+\Delta X(v)$.
In these disks, the rotation period of the molecular
emission region is on the order of 1-10 years. A spectro-astrometric monitoring program should therefore be able to detect changes 
in the azimuth of the asymmetries. 
 
\begin{figure}
\centering
\includegraphics[width=7.0cm]{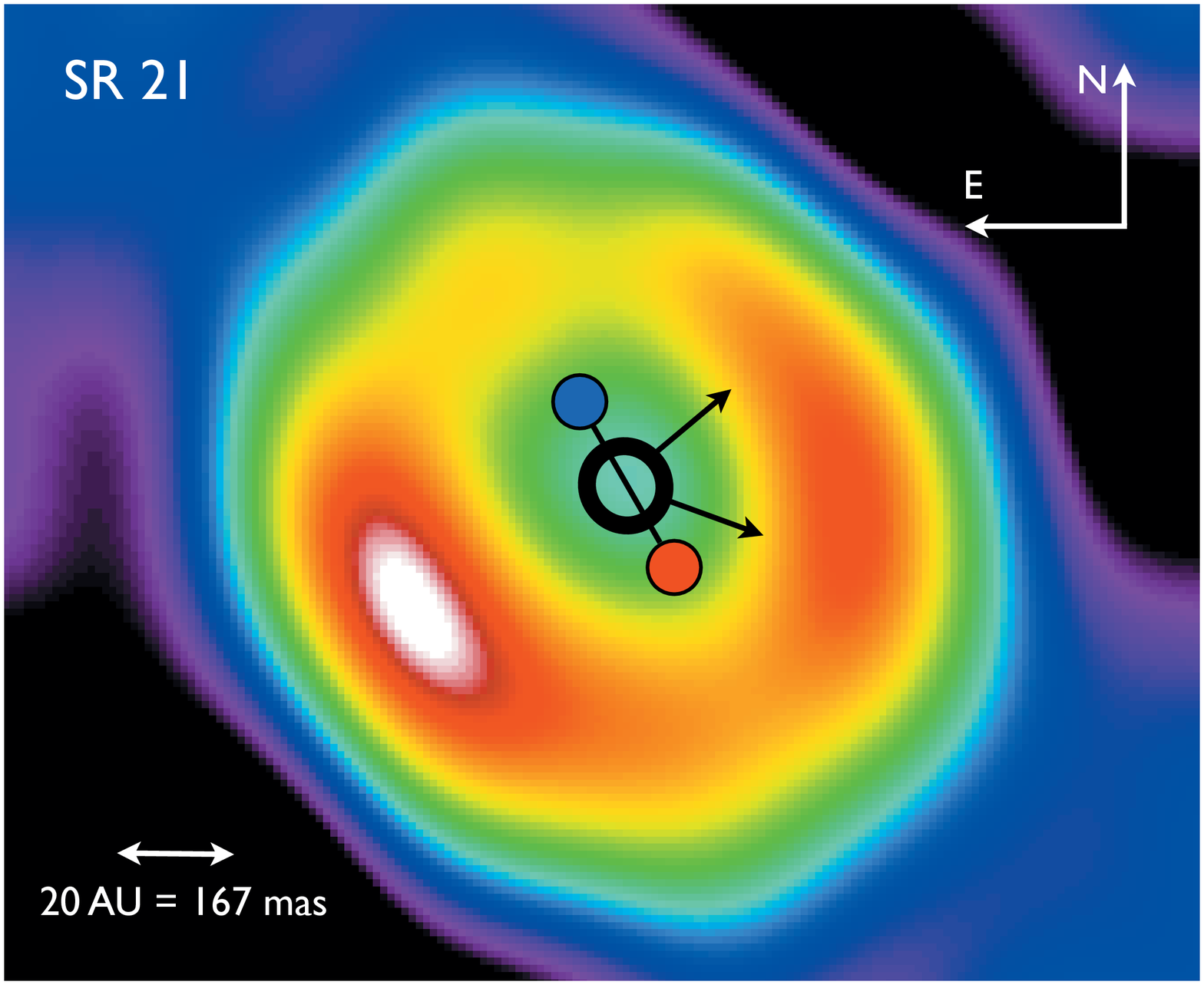}
\includegraphics[width=7.0cm]{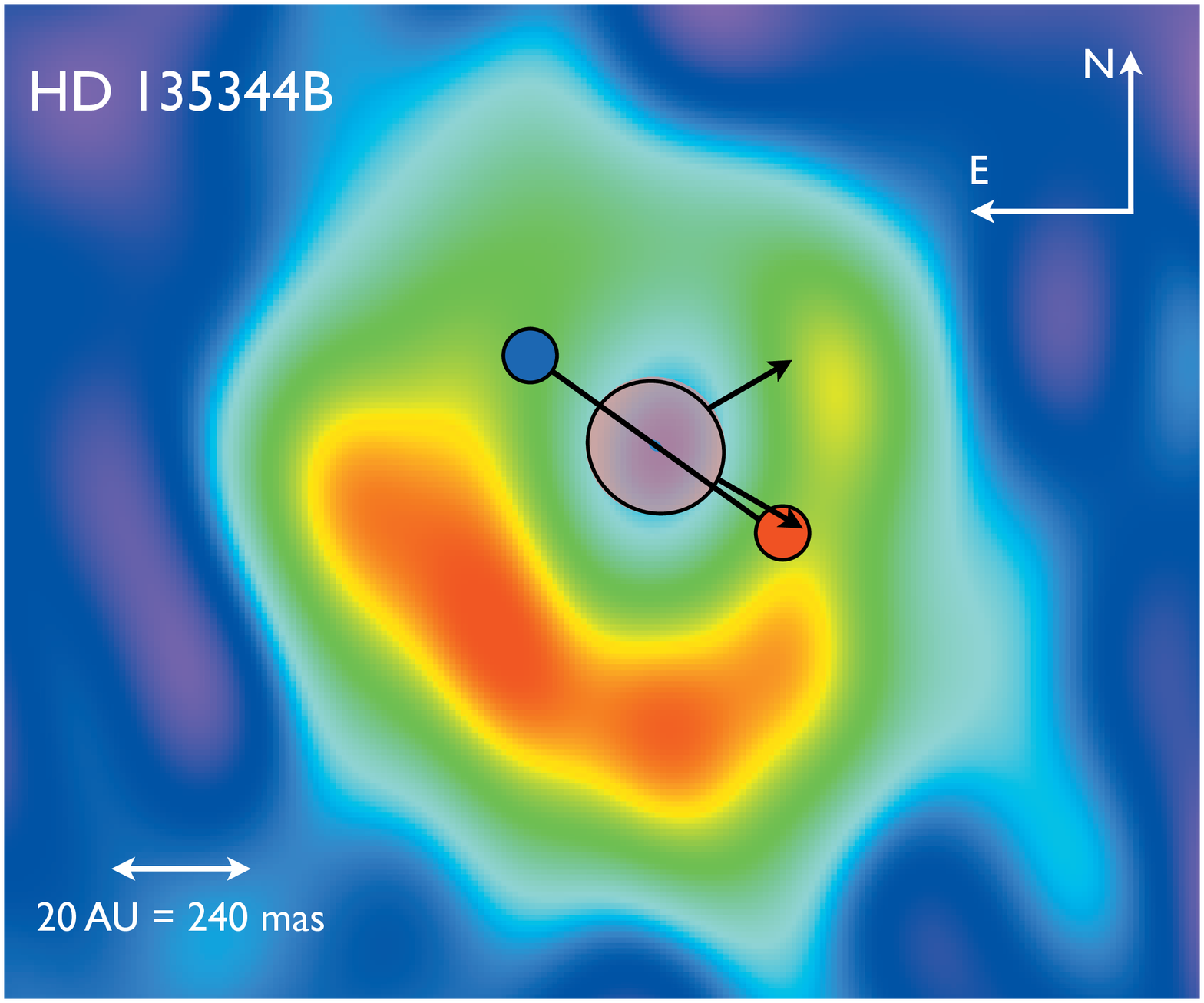}
\caption[]{Sketch of the CO rovibrational line emission relative to the submillimeter continuum emission from \cite{Brownthesis07}
in SR 21 and HD 135344B. The emission in HD135344B extends from 10-15\,AU to within 0.5\,AU from the central
star, whereas the emission from SR 21 is being constrained to a relatively narrow ring around 7\,AU. 
This is illustrated in the figure by a filled region for HD 135344B and a ring
for SR 21. The solid line indicates
the best fit position angle and the ``dumbbell'' shows which side of disk is red-shifted; note that the rotation is measured by
gas on much smaller size scales than the submillimeter continuum emission ring. The arrows indicate the
line asymmetry directions, as defined in the text. }
\label{sketch}
\end{figure}

The distribution of gas in the inner disks of SR 21 and HD 135344B can also be compared to the dust of the disks outside of their inner 
gaps, as imaged by the SMA \citep{Brownthesis07}, clearly showing the presence of the dust gaps inferred from
SED analyses. Fig. \ref{sketch} compares the spectro-astrometric molecular line image information with these SMA images, 
including the asymmetry directions. The best fit position angles as well as the asymmetries appear to 
match the appearance of the dust disks very well, in particular for SR 21. The strongest submillimeter
emission blobs for both SR 21 and HD 135344B are aligned perpendicular to the major axis of the disk, 
indicating that the blobs are not due to a geometric effect in an optically thin disk 
(which would place the blobs parallel to the major axis), but likely indicate real temperature and/or density differences.

\section{Discussion}

We have shown that spectro-astrometry of the fundamental rovibrational
band of CO can directly measure many of the basic 
geometric parameters (position angle, inclination, radial
distribution and departures from axisymmetry) of molecular gas in inner protoplanetary
disks with much fewer ambiguities than simple spectroscopy. 

For the three disks studied in this paper, the high disk masses \citep{Brownthesis07} as well as 
the presence of molecular gas inside the dust gap argue against
photo-evaporation as a gap-forming mechanism. However, is it possible, using the spectro-astrometry, 
to distinguish between grain growth and dynamical clearing? Further, is it possible to
rule out dynamical clearing by a massive, even stellar, companion?
 
\paragraph{SR 21}

The presence of molecular gas in SR 21 at 7\,AU argues against a stellar companion as the cause 
of the observed dust gap at 18\,AU. However, one possible interpretation of the truncation of gas within 7\,AU, is 
the presence of a relatively massive companion at $\lesssim 3.5\,$AU, depending on eccentricity \citep{Artymowicz94}. 
Such a companion, however, should still allow the presence of an inner disk at 0.45\,AU that may 
drain rapidly if not replenished from larger radii, as discussed in \cite{Brown07}. 

Could the dust gap between 7 and 18\,AU be due to grain growth, but with the gas continuing inwards as
a natural extension of the outer disk? By definition, 
dust in protoplanetary disk gaps and holes has much lower optical depth in the infrared in the vertical direction than
disks without such structure. In this case,
infrared rovibrational lines will be much better tracers of the total vertical gas column density than
normally, since the disk midplanes will
no longer be blocked from view by a high optical depth due to dust continuum absorption. 
Indeed, the presence of strong lines of rare isotopologues ($^{13}$CO, C$^{18}$O and C$^{17}$O, see Figure \ref{SingleLines}) 
in emission in SR 21 indicates higher vertical gas column densities than HD 135344B; SR 21
has $^{12}$CO/$^{13}$CO line ratios of $\sim 3$, while HD 135344B has ratios of $\sim 8-10$. 
A high vertical gas column density in SR 21 argues for dust gap formation by grain growth beyond 7\,AU.
A truncation of molecular gas within 7\,AU, however, is more
consistent with dynamical clearing by a massive companion at $\sim 3-4\,$AU. It thus seems
that more than one mechanism is in play in the SR 21 disk. More detailed multi-dimensional radiative
transfer modeling is needed to accurately determine
the vertical CO column densities.

\paragraph{HD 135344B}
 
Our detection of molecular gas extending from within 0.3\,AU of the star, 
to at least 15\,AU argues against clearing by a {\it stellar} companion for HD 135344B, since this
would likely have led to a completely evacuated inner hole. 
Further, the vertical column density of molecular gas seems relatively low, as evidenced by the low  $^{12}$CO/$^{13}$CO line ratios.
Since HD 135344B is surrounded by an otherwise massive disk, the CO rovibrational lines are more consistent with it having
an actual density maximum at the outer dust gap edge, possibly caused by a planetary mass companion orbiting at 10-20\,AU. 
In this case, care must be taken to rule out that photo-dissociation has lowered the
vertical column density of CO, which might happen if grain growth has removed the UV dust opacity. If this
happens, molecular gas will no longer be a good tracer of the total gas column. 
However, even in the absence of opacity-producing dust, 
CO will generally become self shielding, except for very low mass disks \citep{Jonkheid06}.

\paragraph{TW Hya}
Our observed distribution of warm CO gas between 0.11 - 1.5 AU for TW Hya shows
that molecular gas is present in the optically thin dust region found by interferometric dust measurements. It is also consistent with the recent
detection of a young planet at 0.04\,AU, which is likely to truncate the disk at $\sim$0.1\,AU \citep{Setiawan08}. 
The $^{12}$CO/$^{13}$CO line ratio in TW Hya is very high \citep[$\gtrsim 15$, see also][]{Salyk07}, indicating a 
vertical gas column density that is lower than expected from a smooth, monotonic radial surface density profile.
Thus the data indicate that a mechanism other than grain growth is producing the optically thin
inner disk hole. Note that if the 4\,AU inner hole is being dynamically cleared by a planetary companion, it
will not be the companion detected at 0.04\,AU, but another one with an orbit at 1-2\,AU.

Our estimate of the inclination of the inner disk for TW Hya of $4.3\pm 1.0\degr$ brings the TW Hya companion 
mass to $16^{+5}_{-3}\,M_{\rm Jup}$. 
The inner disk inclination, if more representative of the companion orbit inclination than the outer disk inclination of $7\pm 1\degr$
by \cite{Qi04}, places this short period companion in the ``brown dwarf desert'' \citep{Grether06}. Further, the
difference in inclination measured in the inner and outer disks of TW Hya is indicative of a warp in the disk, also consistent
with the presence of a perturbing young planetary system. A warp in a protoplanetary disk can be induced by
an encounter with a passing star, or through dynamical interactions with a planetary system such as has been
suggested for the famous Beta Pictoris debris disk \citep{Golomowski06}. There are no obvious nearby cluster members
to the TW Hya disk, so a stellar encounter is a less likely explanation for a disk warp. A disk warp induced 
by a planet probably requires that the inclination of the orbit of the perturbing planet has been changed relative to
the mean disk inclination by an interaction with another planet. It is therefore tempting to speculate
that there is an additional planet in the TW Hya system, possibly causing the observed 4\,AU gap. 

This paper presents the first results of an extensive spectro-astrometric imaging survey of molecular
gas in the planet forming regions of nearby protoplanetary disks. 
With the powerful resolution of CRIRES of $<1$ milli-arcsec, comparable to or even better than the resolution of
near-infrared interferometers, the CO rovibrational emission from nearby protoplanetary disks 
is generally resolved both spatially and kinematically. A
sensitive interferometric capability at 3-5\,$\micron$ with resolving powers of 50,000-100,000 will be required to improve
on the spectro-astrometry presented here. 
Spatially resolving the emission from other molecules, in particular the water lines 
recently reported in \cite{Salyk08}, 
is likely within the capability of CRIRES, and would directly probe the chemical variations in the planet-forming
region of circumstellar disks.

\acknowledgments{
Support for KMP was provided by NASA through Hubble Fellowship grant \#01201.01 
awarded by the Space Telescope Science Institute, which is operated by the Association of 
Universities for Research in Astronomy, Inc., for NASA, under contract NAS 5-26555.
GAB acknowledges support from the NSF astronomy program.
EvD acknowledges a Netherlands Organization of Scientific Research (NWO) Spinoza Grant.}

\bibliographystyle{apj}
\bibliography{ms}

\end{document}